# Two attacks and counterattacks on the mutual semi-quantum key agreement protocol using Bell states

**Jun Gu[1], Tzonelih Hwang[*]**

*[1]Department of Computer Science and Information Engineering, National Cheng*

*Kung University, No. 1, University Rd., Tainan City, 70101, Taiwan, R.O.C.*

[1] isgujun@163.com

[*]**Corresponding Author:**

Tzonelih Hwang

Distinguished Professor

Department of Computer Science and Information Engineering,

National Cheng Kung University,

No. 1, University Rd.,

Tainan City, 70101, Taiwan, R.O.C.

Email: hwangtl@csie.ncku.edu.tw

TEL: +886-6-2757575 ext. 62524

# Abstract

Recently, a mutual semi-quantum key agreement protocol using Bell states is proposed by Yan et al. (Mod. Phys. Lett. A, 34, 1950294, 2019). The proposed protocol tries to help a quantum participant share a key with a classical participant who just has limited quantum capacities. Yan et al. claimed that both the participants have the same influence on the final shared key. However, this study points out that the classical participant can manipulate the final shared key by himself/herself without being detected. To solve this problem, an improved method is proposed here.

**Keywords** Semi-quantum. Quantum key agreement. Permutation attack. Substitution attack.

## 1. Introduction

Quantum key agreement (QKA) protocol [1] is proposed for helping the involved participants share a fair secret key. Here, 'fair' means none of the proper subsets of the involved participants can determine any part of the final shared key without being detected by the others. In 2004, Zhou et al. [1] proposed a QKA protocol first. Afterword, several QKA protocols [2-7] have been proposed. However, most of these QKA protocols need all the involved participants to have lots of quantum capabilities. To help the participants who just have restricted quantum capacities can be involved in the QKA, the semi-quantum key agreement (SQKA) protocol [8, 9] is proposed.

Recently, Yan et al. [10] proposed a two-party SQKA protocol using Bell states. They claimed that, in their SQKA protocol, both participants have equal contribution to the final shared key. However, this study shows that an involved participant can use two different attack strategies to choose a preferred key as the final shared key without being detected. Hence, to avoid these attacks, a simple modification is proposed here.

The rest of this paper is organized as follows. In Section 2, Yan et al.'s SQKA protocol is reviewed. In Section 3, we show the attacks on Yan et al.'s SQKA protocol and then propose a modified method to avoid them. At last, a conclusion is given in

Section 4.

# 2. A brief review of Yan et al.'s SQKA

Before reviewing Yan et al.'s SQKA protocol [10], some background is introduced first here.

## 2.1 Background

In Yan et al.'s SQKA, four single photons $\{|0\rangle, |1\rangle, |+\rangle = \frac{1}{\sqrt{2}}(|0\rangle + |1\rangle), |-\rangle = \frac{1}{\sqrt{2}}(|0\rangle - |1\rangle)\}$ and four Bell states $\{|\Phi^+\rangle, |\Phi^-\rangle, |\Psi^+\rangle, |\Psi^-\rangle\}$ are used. The details of $\{|\Phi^+\rangle, |\Phi^-\rangle, |\Psi^+\rangle, |\Psi^-\rangle\}$ are described as follows:

$$\begin{cases} |\Phi^+\rangle = \frac{1}{\sqrt{2}}(|00\rangle + |11\rangle) \\ |\Phi^-\rangle = \frac{1}{\sqrt{2}}(|00\rangle - |11\rangle) \\ |\Psi^+\rangle = \frac{1}{\sqrt{2}}(|01\rangle + |10\rangle) \\ |\Psi^-\rangle = \frac{1}{\sqrt{2}}(|01\rangle - |10\rangle) \end{cases} \quad (1)$$

## 2.2 Yan et al.'s SQKA protocol

Suppose that there are two participants Alice and Bob involved in Yan et al.'s SQKA protocol. Alice is a quantum participant who has unrestricted quantum capacities and Bob is a classical participant who is restricted to perform the following four operations.

(a) Generate qubits in Z-basis $\{|0\rangle, |1\rangle\}$.

(b) Measure qubits with Z-basis.

(c) Reorder the qubits via different delay lines.

(d) Send or reflect the qubits.

Then Yan et al.'s SQKA protocol can be described as follows.

**Step 1**: Alice generates $2n$ Bell states $B = \{(q_{A1}, q_{B1}), (q_{A2}, q_{B2}), \cdots, (q_{A2n}, q_{B2n})\}$

where each Bell state is in either $\left|\Phi^+\right\rangle$ or $\left|\Psi^+\right\rangle$ randomly. Then she divides

$B$ into two ordered particle sequences $S_A = \{q_{A1}, q_{A2}, \cdots, q_{A2n}\}$ and

$S_B = \{q_{B1}, q_{B2}, \cdots, q_{B2n}\}$. Subsequently, she sends $S_B$ to Bob.

**Step 2:** Bob generates a random bit sequence $K_B = \{k_{B1}, k_{B2}, \cdots, k_{Bn}\}$. Then, for each qubit received, Bob randomly chooses one of the two following cases.

Case (a). Bob does not perform any operations on this particle.

Case (b). Bob measures this particle with Z-basis and generates a particle

whose value is the XOR of the measurement result and $k_{Bi}$

$(1 \leq i \leq n)$ where $i$ is the $i$ th measured particle. That is, assume

that the measurement result is $\left|r_i\right\rangle \left(r_i \in \{0,1\}\right)$, then Bob generates a new particle in the state $\left|r_i \oplus k_{Bi}\right\rangle$. For example, if the measurement result is $\left|1\right\rangle$ and $k_{Bi}=1$, then the generated qubit will be $\left|0\right\rangle$.

Finally, Bob performs a permutation operation on all the $2n$ particles via different delay lines and then sends them back to Alice.

**Step 3.** After Alice receives the returning qubits, she generates a random $2n$-bit sequence $K_A^r = \{k_{A1}, k_{A2}, \cdots, k_{A2n}\}$ as her secret raw key for further quantum key agreement. Subsequently, she announces $K_A^r$. Then Bob announces the positions of all the particles in Case (a) and the corresponding permutation operation on them.

**Step 4.** Alice uses the published information to check whether there is an eavesdropper

during the qubits transmitted processes or not. That is, for each particle in Case

(a), Alice performs Bell measurement on this particle and its corresponding qubit in $S_A$. Then she checks whether the measurement result is equal to the initial state or not. If the error rate exceeds a predetermined value, this protocol will be aborted. Otherwise, Bob discards the corresponding bits in $K_A^r$ to obtain Alice's secret key $K_A$.

**Step 5**. Bob announces the permutation operation performed on the remaining qubits.

Alice recovers these qubits to a correct order according to this information and then uses Bell measurement to obtain $K_B$. For instance, assume that the initial state of $\{q_{Ai}, q_{Bi}\}$ is $\left| \Phi^+ \right\rangle$ $(\left| \Psi^+ \right\rangle)$. If the measurement result of $\{q_{Ai}, q_{Bi}\}$ is $\left| \Phi^+ \right\rangle$ or $\left| \Phi^- \right\rangle$ $(\left| \Psi^+ \right\rangle$ or $\left| \Psi^- \right\rangle)$, Alice can know $k_{Bi} = 0$. Otherwise,

Alice can know $k_{Bi} = 1$.

**Step 6.** Alice and Bob obtain the final shared key $K_{AB} = K_A \oplus K_B$.

# 3. The loopholes in Yan et al.'s SQKA protocol and an improvement

Yan et al. claimed that neither Alice nor Bob can manipulate the final shared key $K_{AB}$

without being detected. However, this section points out that Bob can use a permutation attack or a substitution attack to choose a preferred final key $K'_{AB}$ instead. Besides, a simple solution is hence proposed.

## 3.1 Permutation attack on Yan et al.'s SQKA protocol

At the end of Step 4, Bob can obtain Alice's secret key $K_A$ and then computes the final shared secret key $K_{AB} = K_A \oplus K_B$. If he does not want to use $K_{AB}$ as the final shared secret key, then he can announce a fake permutation operation in Step 5 instead.

Upon receiving the fake permutation operation, Alice uses a corresponding fake inverse

permutation operation to reorder the remaining qubits and then performs the Bell

measurement to obtain a fake Bob's secret key $K'_B$. At last, she gets a fake final key

$K'_{AB} = K_A \oplus K'_B$ which is chosen by Bob.

For example, assume that $K_A = 1001$, $K_B = 0101$ and there are four particles $\{q_{B1}, q_{B2}, q_{B3}, q_{B4}\}$ performed in Case (b) in Step 2. Moreover, assume that the measurement results of $\{q_{B1}, q_{B2}, q_{B3}, q_{B4}\}$ are $\{|1\rangle_{r_1}, |0\rangle_{r_2}, |0\rangle_{r_3}, |1\rangle_{r_4}\}$, respectively. And the initial states of $\{(q_{A1}, q_{B1}), (q_{A2}, q_{B2}), (q_{A3}, q_{B3}), (q_{A4}, q_{B4})\}$ are $\{|\Phi^+\rangle_{q_{A1}, q_{B1}}, |\Phi^+\rangle_{q_{A2}, q_{B2}}, |\Psi^+\rangle_{q_{A3}, q_{B3}}, |\Psi^+\rangle_{q_{A4}, q_{B4}}\}$, respectively. Under these assumptions, the permutation attack on Yan et al.'s SQKA protocol can be described as follows.

In Step 2, Bob uses Z-basis to measure $\{q_{B1}, q_{B2}, q_{B3}, q_{B4}\}$. Then, he obtains the measurement results $\{|1\rangle_{r_1}, |0\rangle_{r_2}, |0\rangle_{r_3}, |1\rangle_{r_4}\}$ and the qubits $\{q_{A1}, q_{A2}, q_{A3}, q_{A4}\}$ held by Alice collapse into $\{|1\rangle_{q_{A1}}, |0\rangle_{q_{A2}}, |1\rangle_{q_{A3}}, |0\rangle_{q_{A4}}\}$. Subsequently, Bob generates the particles $\{q'_{B1} = |r_1 \oplus k_{B1}\rangle = |1 \oplus 0\rangle = |1\rangle, q'_{B2} = |1\rangle, q'_{B3} = |0\rangle, q'_{B4} = |0\rangle\}$ and performs a permutation operation on them. Finally, he sends these particles back to Alice. At the end of Step 4, Bob obtains the final shared key $K_{AB} = K_A \oplus K_B = 1100$. If Bob wants to choose another key $K'_{AB} = 1010$ instead, in Step 5, he can announce a fake permutation operation with which Alice will recover the $\{q'_{B1}, q'_{B2}, q'_{B3}, q'_{B4}\}$ into $\{q'_{B1}, q'_{B3}, q'_{B2}, q'_{B4}\}$. Then, Alice performs the Bell measurement on $\{(q_{A1}, q'_{B1}), (q_{A2}, q'_{B3}), (q_{A3}, q'_{B2}), (q_{A4}, q'_{B4})\}$ and obtains the measurement results $\{|\Phi^\pm\rangle_{q_{A1}, q'_{B1}}, |\Phi^\pm\rangle_{q_{A2}, q'_{B3}}, |\Phi^\pm\rangle_{q_{A3}, q'_{B2}}, |\Phi^\pm\rangle_{q_{A4}, q'_{B4}}\}$. According to these measurement results and the initial states, Alice gets $K'_B = 0011$ and then obtains a fake final key $K'_{AB} = K_A \oplus K'_B = 1010$ which is manipulated by Bob.

## 3.2 Substitution attack on Yan et al.'s SQKA protocol

In Step 5, if the measurement result obtained by Alice is the same as the initial state, she will think $k_{Bi} = 0$. Hence, in Step 2, Bob can perform Case (a) on the particles where $k_{Bi} = 0$ instead. Then, after Bob obtains $K_A^r$ in Step 3, he can announce arbitrary parts of the particles in Case (a) for eavesdropper detection and the remaining particles will be a substitution for key bits where $k_{Bi} = 0$. This method can help Bob choose a preferred final shared key without being detected.

For example, assume that, $K_A^r = 01000001$ and there are eight particles in Step 2. If Bob performs Case (a) on the particle where $k_{Bi} = 0$, we can assume that the case sequence {a, a, a, b, a, a, b, a} is performed on the eight particles. Here, the case sequence can be converted to a key bit sequence $K_B^r = \{0, 0, 0, 1, 0, 0, 1, 0\} = 00010010$. After Bob gets $K_A^r$, he computes $K_{AB}^r = K_A^r \oplus K_B^r = 01010011$. If Bob wants to use '0000' as the final shared key, he can announce the positions {2, 4, 7, 8} for the eavesdropper detection. Then, after Alice and Bob discard the bits on the positions {2, 4, 7, 8} of $K_{AB}^r = 0\underline{1}0\underline{1}00\underline{11}$, the final shared key will be $K_{AB}^r = 0000$. Similarly, if Bob wants to use '1111' to be the shared key, he can announce the positions {1, 3, 5, 6}.

## 3.3 A solution to the loopholes

In the permutation attack, because the fake permutation operation cannot be detected by Alice, Bob can manipulate the final shared key. If the fake permutation operation can be detected, this problem will be solved. That is, if Bob does not perform any operations on the qubits in Case (b) where $k_{Bi} = 0$, without the correct permutation operation, the Bell measurement performed by Alice on the particles will result in an entanglement swapping [11]. Then the measurement results of the particles where $k_{Bi} = 0$ cannot always be equal to the initial states. Hence, Alice can detect the permutation attack.

In the substitution attack, because Bob obtains $K_A^r$ before announcing the positions for eavesdropper detection, he can exchange parts of key positions with the detection positions to manipulate the final shared key. Hence, if Bob announces the positions before obtaining $K_A^r$, this attack can be solved.

According to the above two methods, the detail of the improvement is as follows.

**Step 1\*** is the same as **Step 1** in Section 2.

**Step 2\***. Bob generates a random bit sequence $K_B = \{k_{B1}, k_{B2}, \cdots, k_{Bn}\}$. Then for each qubit received, Bob randomly chooses one of the two following cases.

Case (a). Bob does not perform any operations on this particle.

Case (b). For the $i$ th $(1 \le i \le n)$ particle in Case (b), if $k_{Bi} = 0$, Bob does not perform any operations on this particle. Otherwise, Bob measures this particle with Z-basis and generates a flipped particle. That is, assume that the measurement result is $|0\rangle$ ($|1\rangle$), then Bob generates a new particle in $|1\rangle$ ($|0\rangle$).

Finally, Bob performs a permutation operation on all the $2n$ particles via different delay lines and then sends them back to Alice.

**Step 3\***. After Alice receives the returning qubits, Bob announces the positions of all the particles in Case (a) and the corresponding permutation operation on them.

**Step 4\***. Similarly, Alice uses the published information to check whether there is an eavesdropper during the qubits transmitted processes or not. If she makes sure that there is not an eavesdropper, she generates a random $n$-bit sequence $K_A = \{k_{A1}, k_{A2}, \cdots, k_{An}\}$ as her secret key and announces $K_A$.

**Step 5\***. Bob announces the permutation operation performed on the remaining qubits. Alice recovers these qubits to a correct order according to this information and then she uses Bell measurement to obtain $K_B$. For instance, assume that the initial state of $\{q_{Ai}, q_{Bi}\}$ is $\left|\Phi^{+}\right\rangle$ ($\left|\Psi^{+}\right\rangle$). If the measurement result of $\{q_{Ai}, q_{Bi}\}$ is $\left|\Phi^{+}\right\rangle$ ($\left|\Psi^{+}\right\rangle$), Alice can know $k_{Bi} = 0$. If the measurement result is $\left|\Psi^{+}\right\rangle$ or $\left|\Psi^{-}\right\rangle$ ($\left|\Phi^{+}\right\rangle$ or $\left|\Phi^{-}\right\rangle$), Alice gets $k_{Bi} = 1$. Moreover, if the measurement result is $\left|\Phi^{-}\right\rangle$ ($\left|\Psi^{-}\right\rangle$), this means there is eavesdropping on it or Bob announced a fake permutation operation.

**Step 6*** is the same as **Step 6** in Section 2.

With this modified method, the problem can be avoided.

# 4. Conclusions

Yan et al. proposed a mutual quantum key agreement protocol using Bell states. However, this study points out that Yan et al.'s SQKA protocol suffers from a permutation attack and a substitution attack. To avoid these attacks, a solution is proposed here.

# Acknowledgment

We would like to thank the Ministry of Science and Technology of the Republic of China, Taiwan for partially supporting this research in finance under the Contract No. MOST 109-2221-E-006-168-; No. MOST 108-2221-E-006-107-.